\documentclass[11pt,a4paper]{article}

\pdfoutput = 1

\usepackage{jcappub}
\usepackage{multirow}
\usepackage{amssymb}

\newcommand{\fett}[1]{\boldsymbol{#1}}

\newcommand{\dd}{{\rm{d}}}
\newcommand{\ii}{{\rm{i}}}

\newcommand{\be}{\begin{equation}}
\newcommand{\ee}{\end{equation}}

\definecolor{darkred}{rgb}{0.5,0,0}
\definecolor{darkgreen}{rgb}{0,0.6,0}
\definecolor{darkblue}{rgb}{0,0,0.5}

\newcommand{\Hc}{\mathcal{H}}

\newcommand{\SONG}{{\sc song}}
\newcommand{\CLASS}{{\sc class}}

\newcommand{\hypF}[4]{\,{_2}F_1 \left( #1, #2, #3, #4 \right)}

\newcommand{\inspire}[1]{\href{http://inspirehep.net/search?p=find+J+#1}
 {{\color{black}[{\color{blue} {\small in}SPIRE}]}}}
\newcommand{\book}[1]{\href{http://inspirehep.net/search?p=#1}
 {{\color{black}[{\color{blue} {\small in}SPIRE}]}}}

\newcommand{\inspired}[1]{\href{http://inspirehep.net/search?p=#1}
 {{\color{black}[{\color{blue} {\small in}SPIRE}]}}}

\newcommand{\nab}{\fett{\nabla}}

\begin{document}

\title{Relativistic initial conditions for N-body simulations}

\date{\today}

\author[a]{Christian Fidler,}
\emailAdd{christian.fidler@uclouvain.be}

\author[b,e]{Thomas Tram,}
\emailAdd{thomas.tram@port.ac.uk}

\author[c,d]{Cornelius Rampf,}
\emailAdd{rampf@thphys.uni-heidelberg.de}

\author[b]{Robert Crittenden,}
\emailAdd{robert.crittenden@port.ac.uk}

\author[b]{Kazuya Koyama}
\emailAdd{kazuya.koyama@port.ac.uk}

\author[b]{and David Wands}
\emailAdd{david.wands@port.ac.uk}

\affiliation[a]{Catholic University of Louvain - Center for Cosmology, Particle Physics and Phenomenology (CP3) 2, Chemin du Cyclotron, B--1348 Louvain-la-Neuve, Belgium}
\affiliation[b]{Institute of Cosmology and Gravitation, University of Portsmouth, Portsmouth PO1 3FX, United Kingdom}
\affiliation[c]{Institut f\"ur Theoretische Physik, Universit\"at Heidelberg, Philosophenweg 16, D--69120 Heidelberg, Germany}
\affiliation[d]{Department of Physics, Israel Institute of Technology --- Technion, Haifa 32000, Israel}
\affiliation[e]{Department of Physics and Astronomy, University of Aarhus, Ny Munkegade 120, DK-8000 Aarhus C, Denmark}

\abstract{
Initial conditions for (Newtonian) cosmological N-body simulations are usually set by re-scaling the present-day power spectrum obtained from linear (relativistic) Boltzmann codes to the desired initial redshift of the simulation. This back-scaling method can account for the effect of inhomogeneous residual thermal radiation at early times, which is absent in the Newtonian simulations. We analyse this procedure from a fully relativistic perspective, employing the recently-proposed Newtonian motion gauge framework. We find that N-body simulations for $\Lambda$CDM cosmology starting from back-scaled initial conditions can be self-consistently embedded in a relativistic space-time with first-order metric potentials calculated using a linear Boltzmann code. This space-time coincides with a simple ``N-body gauge'' for $z<50$ for all observable modes. 
Care must be taken, however, when simulating non-standard cosmologies. As an example, we analyse the back-scaling method in a cosmology with decaying dark matter, and show that metric perturbations become large at early times in the back-scaling approach, indicating a breakdown of the perturbative description. We suggest a suitable "forwards approach" for such cases.
}

\maketitle   

\flushbottom
%%%%%%%%%%%%%%%%%%%%%%%%%%%%%%%%%%%%%%%%%%%%%%%%%%%%%%%%%%
\section{Introduction}
\label{Introduction}

According to the $\Lambda$CDM model, we live in a Universe that is composed of a cosmological constant ($\Lambda$), cold dark matter (CDM), baryons, neutrinos and photons. Today's observed large-scale structure is the consequence of the gravitational evolution of these components from small initial perturbations, and the theory governing the evolution is the coupled set of non-linear Einstein--Boltzmann equations. So far these relativistic equations have been solved not in full non-linearity but by approximate methods within the framework of cosmological perturbation theory (CPT; see e.g.\ \cite{Kodama:1985bj,Malik:2008im,Villa:2015ppa}), utilised by numerical Einstein--Boltzmann solvers. Popular Boltzmann codes that solve the equations to first order include {\sc camb} \cite{Lewis:1999bs} and \CLASS~\cite{Blas:2011rf}, while several codes, including \SONG{}, solve the evolution up to second order \cite{Huang:2012ub,Su:2012gt,Pettinari:2013he}.

These Boltzmann codes determine the gravitational evolution to high accuracy as long as matter and radiation perturbations remain small. This is a good approximation for the first few billion years of the gravitational evolution, but the matter perturbations keep growing during matter domination, making the gravitational collapse of matter non-linear and eventually leading to a complete perturbative breakdown.
To solve for the fully non-linear gravitational collapse, one makes use of cosmological (N-body) simulations \cite{Springel:2005mi,Teyssier:2001cp,Hahn:2015sia} which themselves usually work in the Newtonian approximation (see however \cite{Adamek:2013wja,Adamek:2015eda,Giblin:2015vwq,Hahn:2016roq,Brandbyge:2016raj}), an approach which is justified only on small scales and at sufficiently late times, where the effects of radiation can be neglected. 

To set up initial conditions for such N-body simulations, one has to use perturbative Boltzmann codes and pass the dynamical information about the system over to the Newtonian simulation. 
Perhaps the simplest prescription is to use the Boltzmann codes to calculate the matter density field at high redshift, say at $z=50$, and use this information to set up the initial particle displacements and then evolve these forwards in the N-body simulation. We will refer to this as the ``forwards approach''. However at early times in the standard $\Lambda$CDM cosmology there is residual thermal radiation from the cosmic microwave background and the cosmic neutrino background. Although N-body simulations can include the effect of radiation in the background expansion, they do not describe the effect of inhomogeneous radiation, which tends to suppress the growth of structure. Thus in this naive ``forwards approach'', the large scale density field at the present day ($z=0$) does not match that calculated in perturbative Boltzmann codes which do consistently solve the relativistic evolution to first order.

\begin{figure}[tb]
  \centering
    \includegraphics[width=0.6\textwidth]{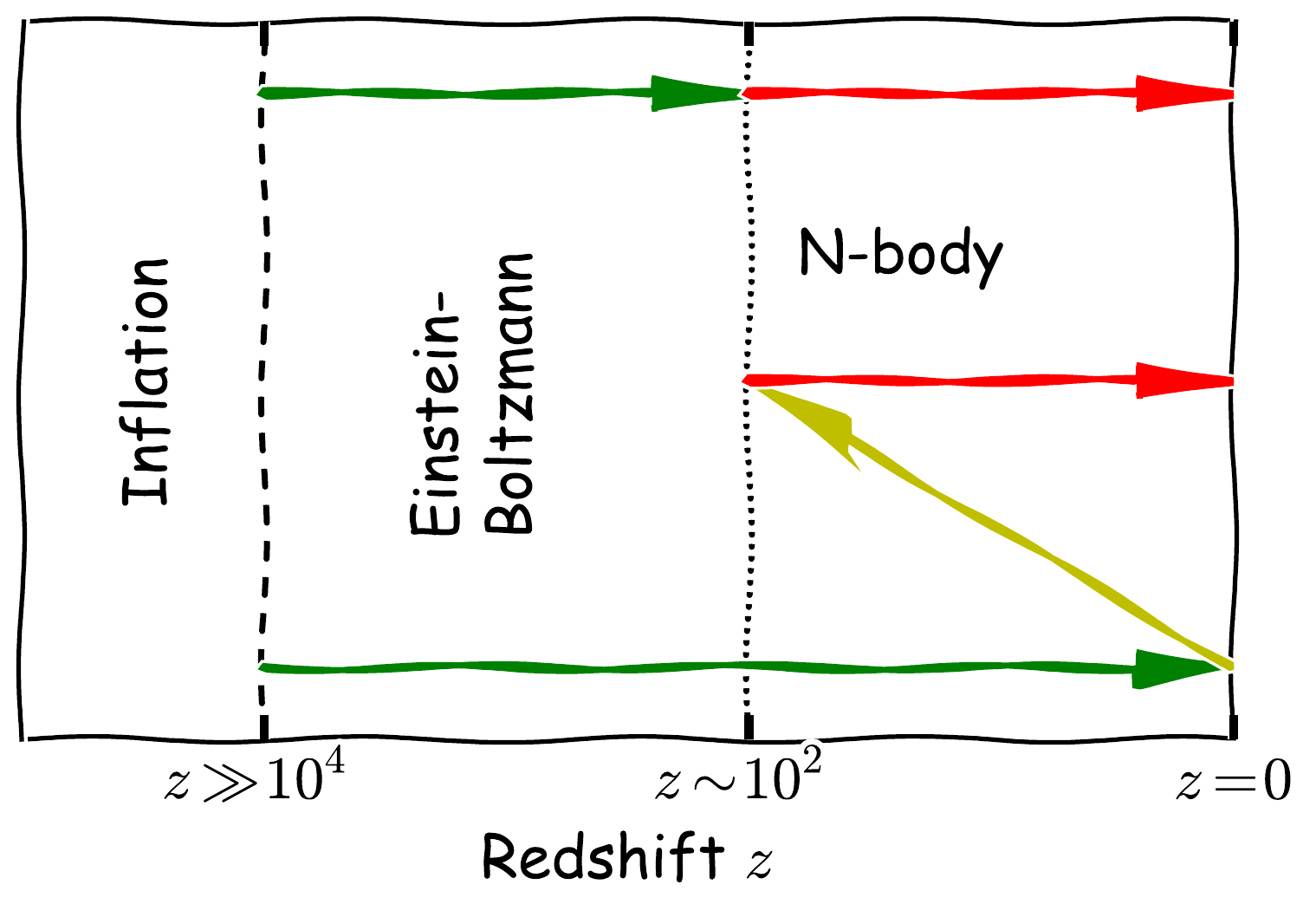}
  \caption{Two different approaches to setting initial conditions for N-body simulations. The ``forwards approach'' (top line) evolves the linear Einstein--Boltzmann perturbations (green arrow) up until a fixed redshift from where the Newtonian N-body solver (red arrow) continues until the present time. In the ``backscaling approach'' (bottom line), the Einstein--Boltzmann system is evolved until $z=0$ (green arrow). The matter perturbations are then scaled back (yellow arrow) using the linear Newtonian theory and are then evolved non-linearly by the N-body simulation (red arrow) to the present day.}
  \label{fig:backscaling}
\end{figure}

A commonly employed work-around is to use a perturbative Boltzmann code to calculate the relativistic density field today and use the linear Newtonian growth factor, which neglects inhomogeneous radiation, to set up an initial density field which, although ``wrong'' at e.g. $z=50$, does enable Newtonian N-body simulations to reproduce the correct linear matter power spectrum on large scales at $z=0$. We refer to this as the ``backscaling approach'', as illustrated in figure~\ref{fig:backscaling}. 
It is the purpose of this paper to analyse whether the methods currently employed in the literature provide a consistent relativistic framework for the analysis of Newtonian N-body simulations, using either forwards or backscaled initial conditions.

This paper is organised as follows. We start by discussing the usual Newtonian framework and the related initial conditions in section~\ref{sec:Newt}. 
In particular we discuss how a ``backscaling approach'' is used to incorporate the consequences of residual radiation perturbations in the initial conditions for Newtonian N-body simulations evolving pressureless dust.
We then briefly summarise the relativistic interpretation in the framework of Newtonian motion gauges in section~\ref{sec:NMrecap}, with the definition of a particular Newtonian motion gauge related to the backscaling method. In section~\ref{sec:rad} we discuss the general solution for the metric in both forwards and backscaled simulations and provide a numerical analysis of the underlying relativistic space-time. We provide an example of a non-standard cosmology with radiation in the late Universe in section~\ref{sec:decay} and conclude in~\ref{sec:conclude}.

\section{Newtonian backscaling}\label{sec:Newt}

Almost all current N-body simulations are performed employing Newtonian gravity. While this is thought to be a good approximation on small scales and at late times, the physics of the early Universe is fundamentally relativistic. This poses a challenge, namely to match relativistic solutions including radiation and matter in the early Universe to Newtonian late-time physics.
In addition, large scales close to the Hubble scale are subject to relativistic effects at all times and relativistic solutions (e.g., for the particle displacements at a given time) require the specification of a gauge, which is absent in a conventional Newtonian simulation.

In the commonly-used ``backscaling approach'',
one first generates the linear matter power spectrum today (at redshift $z=0$) using a first-order Boltzmann code to provide a target for the Newtonian simulation. 
The power spectrum is typically calculated in the synchronous gauge and this sets a relativistic boundary condition for the largest (linear) scales in the simulation at $z=0$, by which time the radiation content of the universe is negligible compared to dark matter and dark energy.

\begin{figure}[tb]
  \centering
    \includegraphics[width=0.7\textwidth]{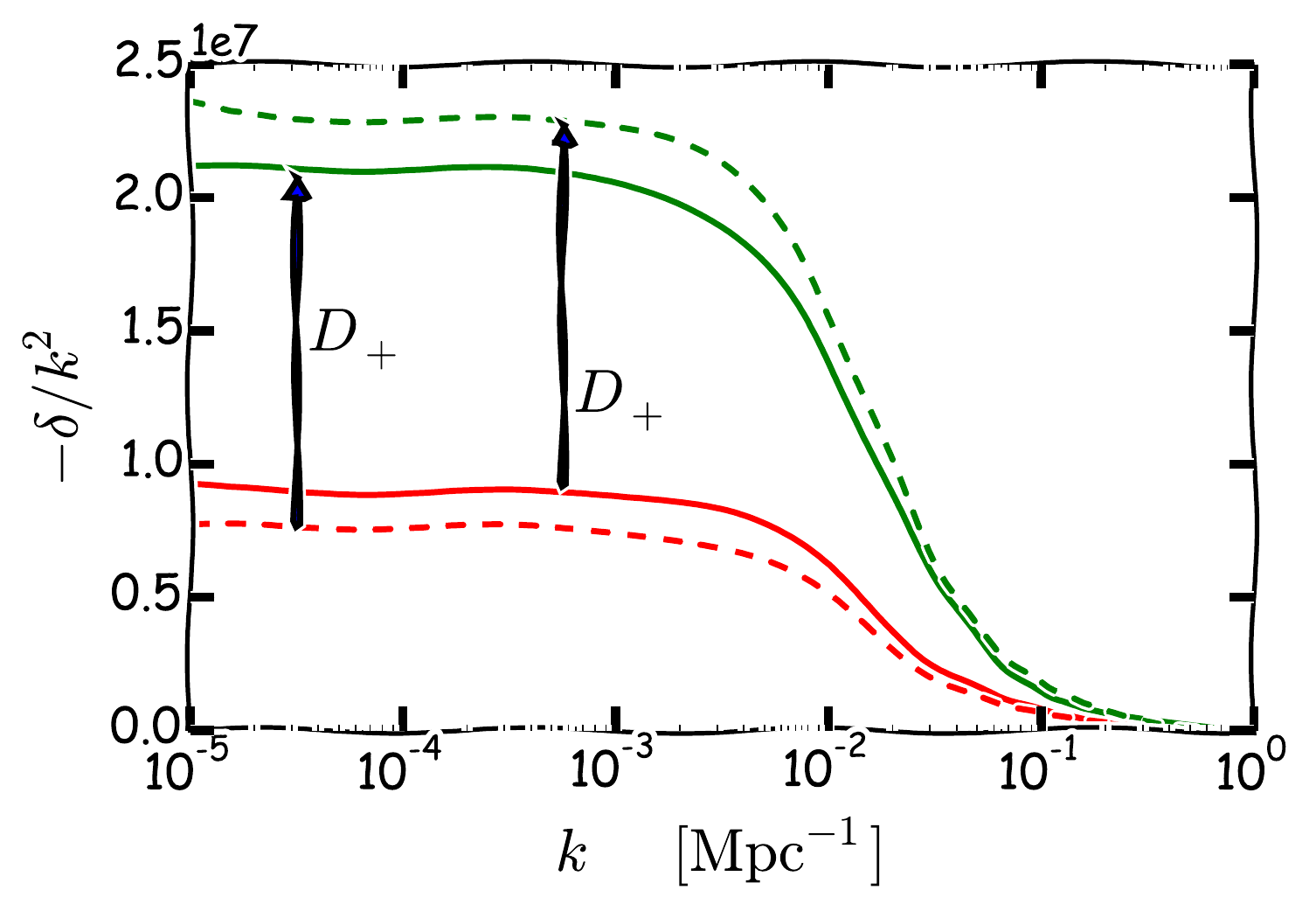}
  \caption{{Comparison between a Newtonian simulation setting initial conditions based on the initial power spectrum for the relativistic density and one using the ``backscaling approach'', employing initial conditions based on the final power spectrum for the relativistic density. The relativistic evolution of the density in synchronous gauge follows the solid lines, from the initial time (red) to the present time (green). 
When evolving using the Newtonian growth function, $D_+$, which neglects the effect of inhomogeneous radiation, the initial relativistic power spectrum (solid red line) leads to the wrong matter power spectrum at the present time (green dashed line). The backwards method, on the other hand, is designed to match the relativistic power spectrum in the linear regime at the present time (solid green line) by starting from a ``fictitious'' initial density field (red dashed line).}}
  \label{fig:figs_KazuyaPlot}
\end{figure}

The amplitude of the matter power spectrum is then scaled back using the \emph{Newtonian} linear growth function for matter perturbations, $D_+$, to set initial conditions when the nonlinear, but Newtonian N-body simulation is initialised, as shown in figure~\ref{fig:figs_KazuyaPlot}.  Matter-only simulations initialised in this way will end up, by construction, with the correct final matter power spectrum on the largest scales where the N-body simulation reproduces the Newtonian linear growth function.
Instead of performing a simulation of the actual Universe, backscaling can be understood as simulating an artificial radiation-free Universe that is designed to mimic our Universe on large scales and at the present time. On the other hand, small scales are assumed to be well-described in the Newtonian theory and they should remain unaffected as long as there is no significant transfer to small scale physics from large scales, where the actual relativistic solution has been replaced by the Newtonian backscaling solution.  

The linearised Newtonian equations of motion employed for backscaling correspond to the Poisson, continuity and Euler equations for the Newtonian potential, $\Phi^{\rm N}$, the Newtonian density contrast, $\delta_{\rm cdm}^{\rm N}$, and the Newtonian velocity, $v_{\rm cdm}^{\rm N}$: 
\begin{subequations}\label{eq:Newton}
\begin{align}  
	 k^2 \Phi^{\rm N} &= 4\pi G a^2 \bar{\rho}_{\rm{cdm}}^{\rm N}  \delta_{\rm cdm}^{\rm N} \,, \label{NPoisson}\\
	 \dot{\delta}_{\rm cdm}^{\rm N} + k v_{\rm cdm}^{\rm N} &= 0 \,, \\ 
   \left[ \partial_\tau + {\cal H} \right] v_{\rm cdm}^{\rm N} &= -k \Phi^{\rm N} \label{NEuler}\,,
\end{align} 
\end{subequations}
These three equations can be combined into the well-known ordinary differential equation for the Newtonian growth of structure in an expanding cosmology with scale factor $a(\tau)$:
\be
\label{eq:Newtoniangrowth}
\left(\partial_{\tau} +\mathcal{H}\right)\dot{\delta}_{\rm cdm}^{\rm N} = 4 \pi G a^2 \bar \rho_{\rm cdm}\delta_{\rm cdm}^{\rm N} \,.
\ee
We use the conformal time $\tau$ defined by $a \dd \tau = \dd t$, where $t$ is the cosmic time. We denote partial derivatives with respect to conformal time with $\partial_\tau$ or an overdot, and the conformal Hubble rate ${\cal H} \equiv \dot a/a$.

Being a second-order differential equation, it has two linearly-independent solutions and we label these as the growing mode, $D_{+}$, and the decaying mode, $D_{-}$. We normalise both to unity at the present time: $D_{\pm}(z=0) = 1$.
The general solution can be given in terms of these two modes as 
\be\label{eq:decNew}
\delta^{\rm N}(\tau) = C_{+}^{\delta^{\rm N}} D_+(\tau) + C_{-}^{\delta^{\rm N}} D_-(\tau) \,,
\ee
where $C_{\pm}^{\delta^{\rm N}}$ are the remaining integration constants fixed by the boundary conditions (typically by the initial displacement field in the particle simulation).
The Wronskian of the two modes is defined by
\be
\label{Wronskian}
W = D_+ \dot{D}_- - D_- \dot{D}_+ .
\ee
It has the particularly simple equation of motion $\dot{W} = -\Hc W \,,$
and is thus required to decay as $a^{-1}$. This provides a simple but powerful numerical test for the linear independence of the two modes. 

Backscaling, using a pure growing mode, from the present relativistic synchronous gauge density, $\delta^{\rm syn}(z=0)$, thus corresponds to setting the initial density contrast in the Newtonian simulation
\be \label{eq:backIcs}
\delta^{\rm N}(\tau_{\rm ini}) 
= D_{+}(\tau_{\rm ini}) \,\delta^{\rm syn}(z = 0) \,.
\ee
In the absence of radiation an analytic solution for the growing mode exists \cite{Demianski:2005is}
\be \label{eq:Newgrowth2}
D^{\text{analytic}}_+ 
\propto  \hypF{\frac{1}{3}}{1}{\frac{11}{6}}{-\frac{\Omega_\Lambda}{\Omega_{\rm m}}a^3} \,,
\ee
where 
${}_2F_1$ is Gauss' hypergeometric function. In the presence of radiation, equation~\eqref{eq:Newgrowth2} no longer holds. However, one may approximate $D_+$ by the integral
\be \label{eq:Newgrowth1}
D^\text{approx}_+ 
\propto \frac{\Hc}{a} \int \dd a'\, \Hc^{-3}(a') \,,
\ee
including the energy density of radiation in the Hubble expansion $\Hc$.\footnote{This quantity was computed (but not used internally) by \CLASS{} \textsc{2.5.1} where it was denoted ``\texttt{gr.fac.\ D}''. \CLASS{} \textsc{2.6.0} now computes $D_+$ and the growth factor $f$ correctly by directly solving the differential equation.} 

\begin{figure}[tb]
  \centering
    \includegraphics[width=.95\textwidth]{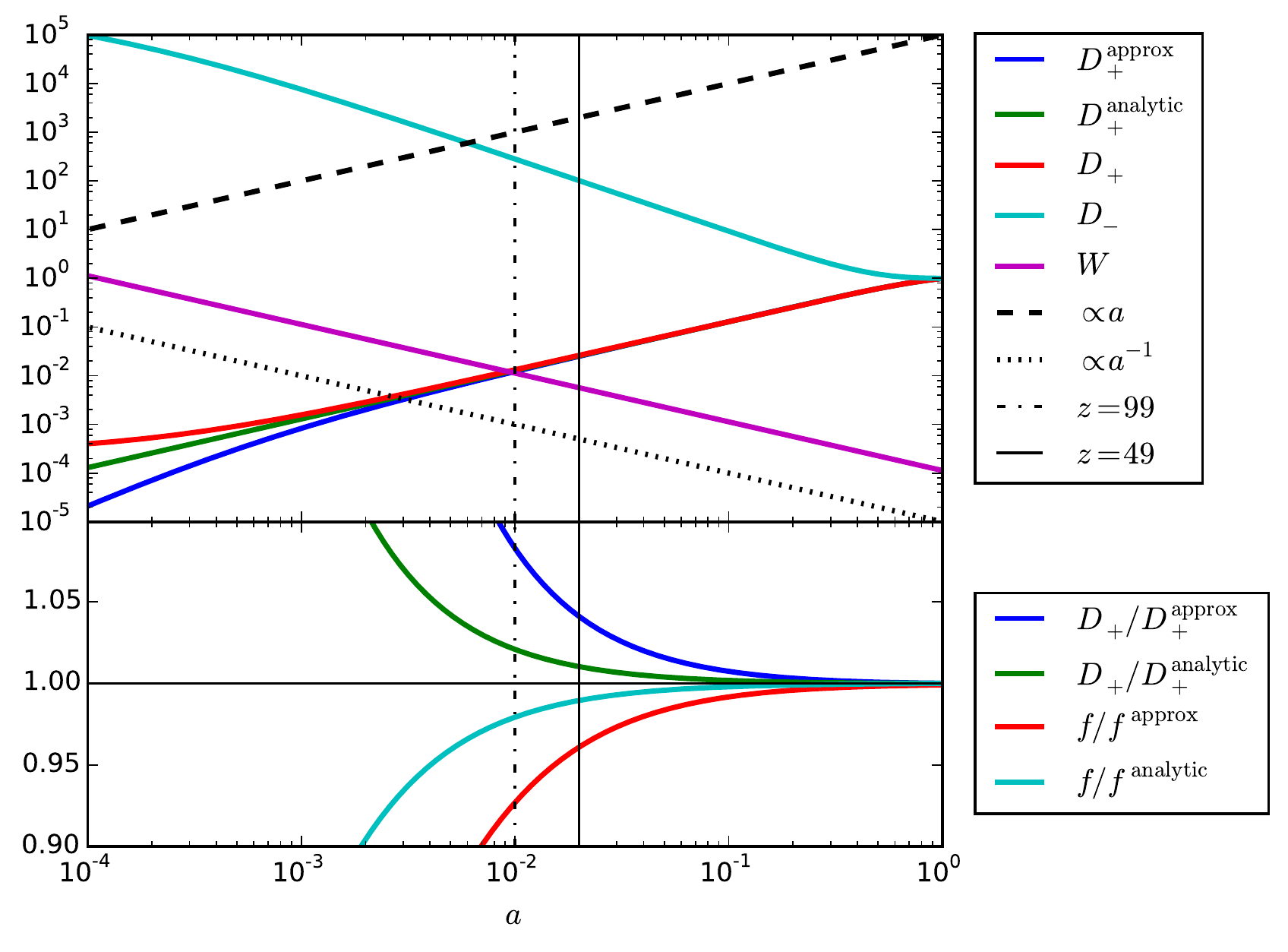}
  \caption{The Newtonian growing and decaying modes, $D_+$ and $D_-$ in \eqref{eq:decNew}, calculated in a standard $\Lambda$CDM cosmology with massless neutrinos $\left(H_0=67.6\frac{\text{km}}{s \text{Mpc}},\Omega_m=0.29,\Omega_\Lambda=0.71 \right)$. We show for comparison the integral approximation, $D^\text{approx}_+$ in \eqref{eq:Newgrowth1}, and the analytic growth factor, $D^\text{analytic}_+$ in \eqref{eq:Newgrowth2}, for the same cosmology. The error is at the \% level at $z=49$, and the same holds for the logarithmic derivative, $f\equiv {\dd\log D_+}/{\dd\log a}$. 
  The Wronskian $W$ of $D_+$ and $D_-$ is proportional to $a^{-1}$ and shows that our two modes are linearly independent.}
   \label{fig:figs_GrowingDecayingBackground}
\end{figure}

Our numerical computations of $D_+$ and $D_-$ are shown in figure~\ref{fig:figs_GrowingDecayingBackground}. This demonstrates that $D^{\rm approx}_+$, $D^\text{analytic}_+$ and $D_+$ disagree at the percent level before redshift $z=49$. The same holds for the logarithmic derivative, $f\equiv {\dd\log D_+}/{\dd\log a}$, used to set the initial velocity field. The analytic expression~\eqref{eq:Newgrowth2} works better than the approximate form~\eqref{eq:Newgrowth1}, but even a 1\% error in the initial power spectrum is unacceptable and one should solve the second order differential equation \eqref{eq:Newtoniangrowth} directly when radiation cannot be neglected. 

In a numerical simulation one needs to set up particles to represent the density $\delta^{\rm N}$ and the three velocity $v^{{\rm N},i}$ in a real space. In practice, the density field is constructed by displacing initially homogenous N-body particles according to the Zel'dovich displacement field 
\be
\nab^i \Psi_i(\tau_{\rm ini}) =- \delta^{\rm N}(\tau_{\rm ini}) 
= - D_{+}(\tau_{\rm ini}) \,\delta^{\rm syn}(z = 0)\,.
\ee
Assuming a pure growing mode, the velocities may be obtained directly from the density: $\nab^i v_i^{\rm N} = -\frac{\dot{D}_+}{\Hc D_+} \delta^{\rm N}$. Employing the displacement field we find
\be
v^{\rm N}_i(\tau_{\rm ini}) = f(\tau_{\rm ini}) \Psi_i(\tau_{\rm ini})\,.
\ee  
This allows for an efficient way to generate both, the initial velocities and the initial particle distribution, based only on the displacement field.

\section{The Newtonian motion gauge framework} \label{sec:NMrecap}

The conventional Newtonian backscaling lacks a consistent relativistic interpretation to quantify the underlying approximations and limitations. 
This paper aims to solve these open issues and provide a fully relativistic framework, making use of the recently introduced Newtonian motion gauges~\cite{Fidler:2016tir}.

The idea of the Newtonian motion (Nm) approach is to choose a gauge in which, by construction, the CDM component 
follows Newtonian trajectories, whilst incorporating relativistic effects in the underlying space-time. The relativistic trajectories 
will then coincide with the Newtonian ones, which implies that
we can identify a relativistic space-time complementing a Newtonian simulation.
The challenge of relativistic particle evolution is thus separated into a purely Newtonian, but highly non-linear part and a relativistic, but perturbative part. Particle trajectories can be solved by conventional Newtonian N-body simulations, while a standard linear Einstein--Boltzmann code is used for the computation of the underlying relativistic space-time.  

\subsection{Notation and conventions}

We adopt the same conventions as ref.~\cite{Fidler:2016tir}, and use
the following Fourier metric perturbations about a homogeneous and isotropic Friedmann--Lema\^itre--Robertson--Walker background,
\begin{subequations}
\label{metric-potentials}
\begin{align}
  g_{00} &= -a^2 \left( 1 + 2A \right) \,, \\
  g_{0i} &=  a^2\, \ii \hat k_i B \,, \\
  g_{ij} &= a^2 \left[ \delta_{ij} \left( 1 + 2 H_{\rm L} \right) + 2 \left( \delta_{ij}/3 - \hat k_i \hat k_j \right) H_{\rm T} \right] \,,
\end{align}
where $\hat k_i \equiv k_i/k$, with $k \equiv |\fett{k}|$,
 $A$ is the perturbation of the lapse function, $B$ is a scalar perturbation in the shift, and $H_{\rm L}$ and $H_{\rm T}$ are respectively the trace and trace-free scalar perturbations of the spatial metric. 
The matter and radiation content is characterised by the stress-energy tensor 
\begin{align} 
	T^{0}_{\phantom{0}0} &= - \sum_\alpha \bar \rho_\alpha \left( 1  + \delta_\alpha \right) \equiv - \bar \rho \left( 1  + \delta \right) \,, \\
	T_{{\phantom{0}}0}^i &=  \sum_\alpha (\bar \rho_\alpha +\bar p_\alpha ) \,\ii \hat k^i v_\alpha \equiv (\bar \rho +\bar p )\, \ii \hat k^i v \,, \\ 
	T^{i}_{\phantom{i}j} &= \sum_\alpha (\bar p_\alpha+\delta p_\alpha ) \delta^i_j + \bar p_\alpha \left( \delta_j^i/3 - \hat k^i \hat k_j \right) \Pi_\alpha 
  \equiv (\bar p+\delta p ) \delta^i_j + \bar p \left( \delta_j^i/3 - \hat k^i \hat k_j \right)  \Pi  \,, \label{Tmunu}
\end{align}
\end{subequations}
where the summation over $\alpha$ runs over all fluid species, i.e.,
CDM, baryons, photons and neutrinos; $\delta \equiv (\rho - \bar \rho) / \bar \rho$ 
is the density contrast and $\bar \rho$ the background density,
$\overline p$ the background pressure and its perturbation $\delta p$, and $\Pi$ is the anisotropic stress of radiation. 

\subsection{Relativistic dynamics}

The Einstein--Boltzmann equations provide the following equations of motion for the CDM component to first order in a yet unfixed gauge \cite{Hu:2004xd},
\begin{subequations}
\begin{align} 
 	 k^2 \Phi &= 4 \pi G a^2 \left[ \bar \rho \delta + 3 {\cal H}  \left( \bar \rho+ \bar p \right)  ( v - B ) /k \right]  \,, \label{eq:EE2}\\
\dot{\delta}_{\rm{cdm}} + k v_{\rm cdm} &= - 3 \dot H_{\rm L} \,, \label{eq:NMconti} \\  
 \left[ \partial_\tau + {\cal H} \right] v_{\rm cdm} &= -k ( \Phi +  \gamma ) \,, \label{eq:NMvelo}  
\end{align}
\end{subequations}
where $\Phi$ is the gauge-invariant Bardeen potential \cite{Bardeen:1980kt}
and $\gamma$ is a relativistic correction describing the forces acting on the dark matter particles, 
\be \label{def:gammaFourier}
 - k^2  \gamma \equiv  \left( \partial_\tau + {\cal H} \right) \dot{ H}_{\rm T} - 8\pi G a^2 \bar p \Pi \,.
\ee
The Newtonian motion gauges are defined by the spatial gauge condition~\cite{Fidler:2016tir}
\be \label{eq:NMgauge}
\gamma + \Phi= \Phi^{\rm N} \, ,
\ee
which guarantees a Newtonian motion for the dark matter particles. That is, the relativistic Euler equation~(\ref{eq:NMvelo}) has the same form as the Newtonian Euler equation (\ref{NEuler}), where the Newtonian potential, $\Phi^{\rm N}$, is given by the Newtonian Poisson equation~(\ref{NPoisson}). Note that the Newtonian density $\delta^{\rm N}_{\rm cdm}$ is the coordinate density and does not include the relativistic volume deformation of the underlying space-time; it is related to the full (relativistic) matter density, $\delta_{\rm cdm}$, via
\be \label{eq:gemoetry}
\delta^{\rm N}_{\rm cdm} = \delta_{\rm cdm} + 3 H_{\rm L}.
\ee
Thus, although the particle displacements in a Newtonian simulation now agree with the relativistic particle displacements in a Newtonian motion gauge, the Newtonian density calculated in an N-body simulation does not in general match the relativistic density.

Note that the temporal gauge is not fixed by the Newtonian motion gauge and we leave it unspecified for now.

\subsection{Relation to the N-body gauge}
\label{forwards}

Another useful and related gauge is the N-body gauge~\cite{Fidler:2015npa} (denoted in the following equations by the ``Nb'' superscript). It is defined with a gauge condition such that $H_{\rm L}^{\rm Nb} = 0$, so that the physical volumes are not modified by relativistic volume deformations. As a result the relativistic matter density corresponds to the simulation density computed by a naive counting of particles. 

The temporal gauge condition for the N-body gauge is $B^{\rm Nb} = v^{\rm Nb} $, so that the constant time hyper-surfaces are orthogonal to the total matter and radiation 4-velocity. As a result, the N-body gauge further satisfies the spatial constraint
$H_{\rm T}^{\rm Nb} = 3\zeta$, 
where $\zeta$ is the
gauge-independent 
comoving curvature perturbation~\cite{Hu:2004xd}
\be
\zeta = H_{\rm L} +\frac{1}{3} H_{\rm T} + \Hc k^{-1} (B - v) \,. 
\ee

We have previously shown in ref.~\cite{Fidler:2016tir} that the Newtonian motion gauge can be completely fixed by identifying it with the N-body gauge at a particular time. In that work, we focus on matching with the N-body gauge at an early epoch, and evolving forward to see how the gauge choices diverge. In the absence of radiation, the two gauges remain synchronised, but they quickly diverge if the matching occurs sufficiently early when radiation is important. We refer to this type of gauge fixing as \emph{forwards Newtonian motion gauges.} 

\subsection{Backscaling and Newtonian motion gauges}

Instead of connecting the Newtonian motion gauges to the N-body gauge at the initial time, we may alternatively match the Newtonian motion gauge to the N-body gauge at the present time. As a Newtonian simulation is interpreted in the corresponding Newtonian motion gauge, this implies that the output of the Newtonian simulation must match the present day power spectrum in the N-body gauge on the linear scales.

The Newtonian motion gauge and N-body gauge remain synchronised in the absence of radiation; 
working backwards in time from the present, this correspondence holds 
in the standard $\Lambda$CDM cosmology until the beginning of the matter era at which time 
radiation becomes important and the two gauges begin to diverge. 
Therefore, the initial conditions in this Newtonian motion gauge no longer match the N-body gauge initial power spectrum in the presence of radiation.

We call this type of Newtonian motion gauge the \emph{backwards Newtonian motion gauges}. In the following we show that they provide a relativistic embedding of the usual Newtonian ``backscaling approach''.  

\section{Solutions to the Newtonian motion gauge condition}\label{sec:rad}

The Newtonian motion gauge condition (eq.\,\ref{eq:NMgauge}) is equivalent to a second-order differential equation for $H_{\rm T}$, as can be seen from the explicit form of $\gamma$ in eq.\,(\ref{def:gammaFourier}).
Without specifying the cosmology, the metric potential $H_{\rm T}$ evolves according to:
\be\label{eq:metricfull}
\left(\partial_{\tau} +\mathcal{H}\right)\dot{H}_{\rm T} - 4 \pi G a^2 \bar \rho_{\rm cdm}\left( H_{\rm T} -3\zeta \right)= S
\ee
with a source term $S$ that is non-zero in the presence of radiation,
\be
S = 4 \pi G a^2 \left(\bar \rho_{\rm other}\delta_{\rm other} +  3 \mathcal{H}k^{-1} (\bar \rho+\bar p)_{\rm other}(v-B) + 2 \bar p \Pi\right) \,,
\ee 
where $\bar{\rho}_{\rm other}$, etc, represent fluid species other than matter.

In the radiation-free limit, we can ignore the source term to find 
\be \label{eq:Metricgrowth}
\left(\partial_{\tau} +\mathcal{H}\right)\dot{H}_{\rm T} = 4 \pi G a^2 \bar \rho_{\rm cdm}\left( H_{\rm T} -3\zeta \right).
\ee
The homogenous part of the differential equation for $H_T$ (eq.\,\ref{eq:Metricgrowth}) is identical to the equation for the Newtonian density (eq.\,\ref{eq:Newtoniangrowth}). It therefore shares the same linearly independent solutions $D_{+}(\tau)$ and $D_-(\tau)$. The particular solution is driven by the term proportional to $\zeta$ on the right-hand side.
Since $\zeta$ is constant in the absence of radiation, we find the simple particular solution $H_{\rm T} = 3 \zeta$, and
the full solution is
\begin{eqnarray}\label{eq:metricdeco}
H_{\rm T}(\tau) &=& C_{+}^{H_{\rm T}} D_+(\tau) + C_{-}^{H_{\rm T}} D_-(\tau) + 3 \zeta, \label{eq:Newtondeco}
\end{eqnarray}
where the integration constants $C_{+}^{H_{\rm T}}$ and $C_{-}^{H_{\rm T}}$ encode the two parameter residual gauge freedom of the Newtonian motion gauges.
The spatial gauge condition of the N-body gauge is equivalent to $H_{\rm T}(\tau) = 3 \zeta$ and thus setting $C_{+}^{H_{\rm T}} = C_{-}^{H_{\rm T}} = 0$ corresponds to fixing the Newtonian motion gauge to the N-body gauge.\footnote{We remark that it is further possible to set $C_{+}^{H_{\rm T}} = -C_{-}^{H_{\rm T}}\neq0$, but this choice is only equal to the N-body gauge momentarily at the present time and rapidly diverges to a different gauge. Mathematically it would correspond to backscaling methods including decaying modes.} 

However, in the early Universe radiation plays an important role and we must solve the full equation~(\ref{eq:metricfull}).
Using the method of variation of constants, a solution of the space-time can be constructed from equation~(\ref{eq:metricdeco}) by adding a time dependence to the coefficients, 
\be
H_{\rm T} = C_{+}^{H_{\rm T}}(\tau) D_+(\tau) + C_{-}^{H_{\rm T}}(\tau) D_-(\tau) + 3 \zeta \,.
\ee
For this ansatz to solve equation~(\ref{eq:metricfull}) the coefficients must fulfil the conditions
\begin{align}
	 \dot{C}_{+}^{H_{\rm T}} D_+ + \dot{C}_{-}^{H_{\rm T}} D_- &= 0 \,,\\
	 \dot{C}_{+}^{H_{\rm T}} \dot{D}_+ + \dot{C}_{-}^{H_{\rm T}}\dot{D}_- &= S - 3\ddot{\zeta} - 3 \mathcal{H}\dot{\zeta} \equiv \tilde{S} \,,
\end{align}
where the corrections included in $\tilde{S}$ reflect that the comoving curvature is no longer conserved in the presence of relativistic species. 

In the backwards Newtonian motion gauge, the equations above can be integrated with the remaining integration constants fixed by the specification of the gauge at the final time: 
$C_{+}^{H_{\rm T}}(\tau_{\rm final}) = C_{-}^{H_{\rm T}}(\tau_{\rm final}) = 0$~(corresponding to the N-body gauge).
Using the Wronskian $W$, defined in \eqref{Wronskian}, the time-dependent coefficients can be found to be
\be
	C_{\pm}^{H_{\rm T}}(\tau) = \pm \int \limits_{\tau}^{\tau_{\rm final}} \,\tilde{S}(\tilde \tau) D_\mp(\tilde \tau) \,W(\tilde \tau)^{-1} \dd \tilde \tau \,. 
\ee
These coefficients are zero as long as radiation does not have an impact after time $\tau$. This means that even including radiation at early times we will have $H_{\rm T} = 3\zeta = \text{const.}$ for most of the cosmic evolution. Only at early times will the gauge depart from N-body gauge due to the impact of radiation.

\begin{figure}[tb]
  \centering
    \includegraphics[width=\textwidth]{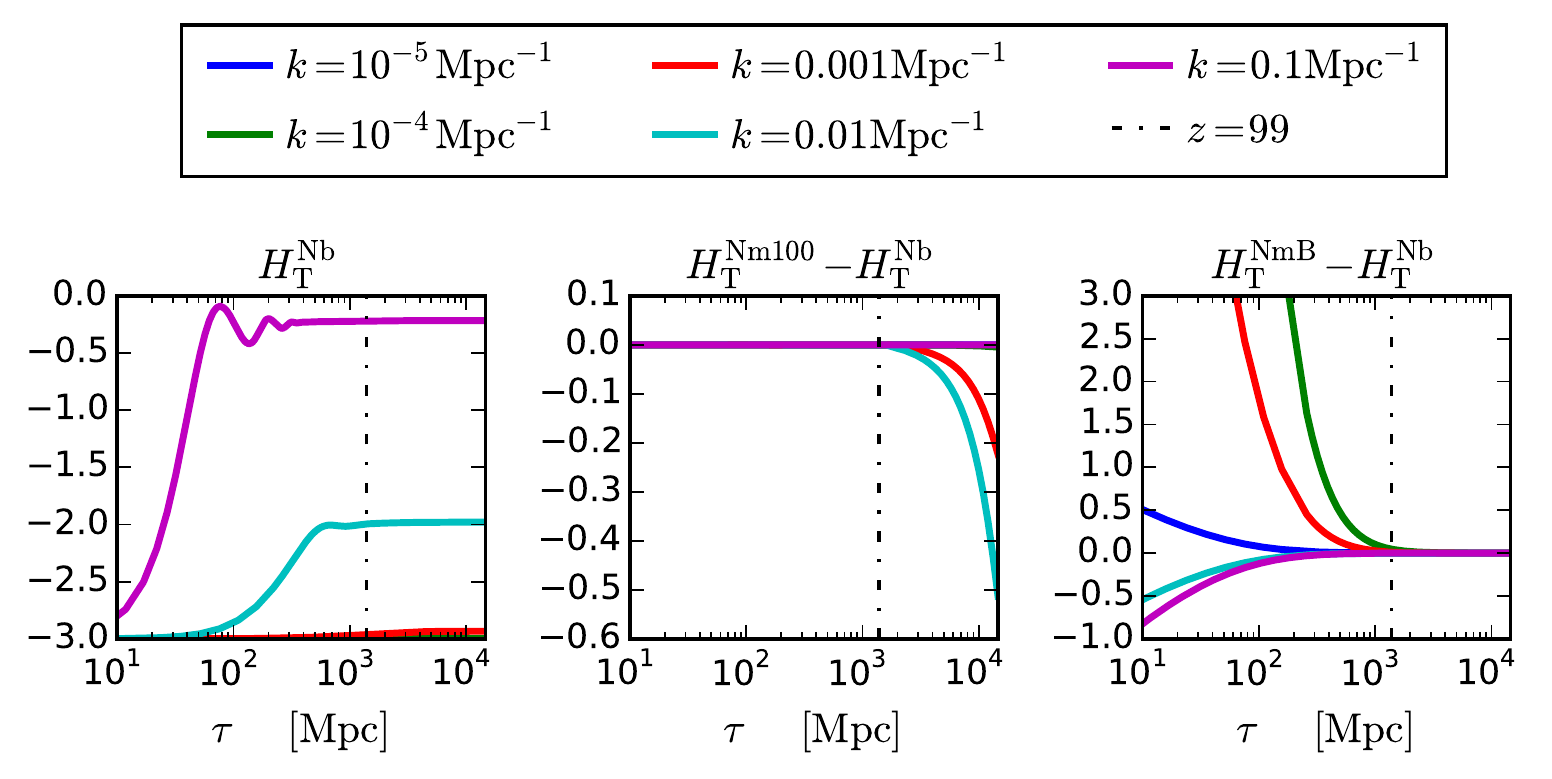}
  \caption{Comparison of the metric potentials, $H_{\rm T}$, in the backward and the forward Newtonian motion gauge approach in a standard $\Lambda$CDM cosmology. \emph{Left panel:} $H_{\rm T}^\text{Nb}=3\zeta$ in the N-body gauge. \emph{Centre panel:} The deviation from the N-body gauge metric potential of $H_{\rm T}^{\text{Nm}100}$ in the ``forwards approach'' when the Newtonian motion gauge is initiated at $z=100$. The modes starts to deviate due to the residual radiation. \emph{Right panel:} The deviation from the N-body gauge metric potential of $H_{\rm T}^{\text{NmB}}$ in the ``backscaling approach''. All modes initially differ from the N-body gauge potential but they are quickly pushed towards the N-body gauge due to the residual radiation. The modes are normalised to $\zeta=-1$ at super-horizon scales according to the \CLASS{}-convention.}
  \label{fig:figs_H_T_decaying2}
\end{figure}

\begin{figure}[tb]
  \centering
    \includegraphics[width=\textwidth]{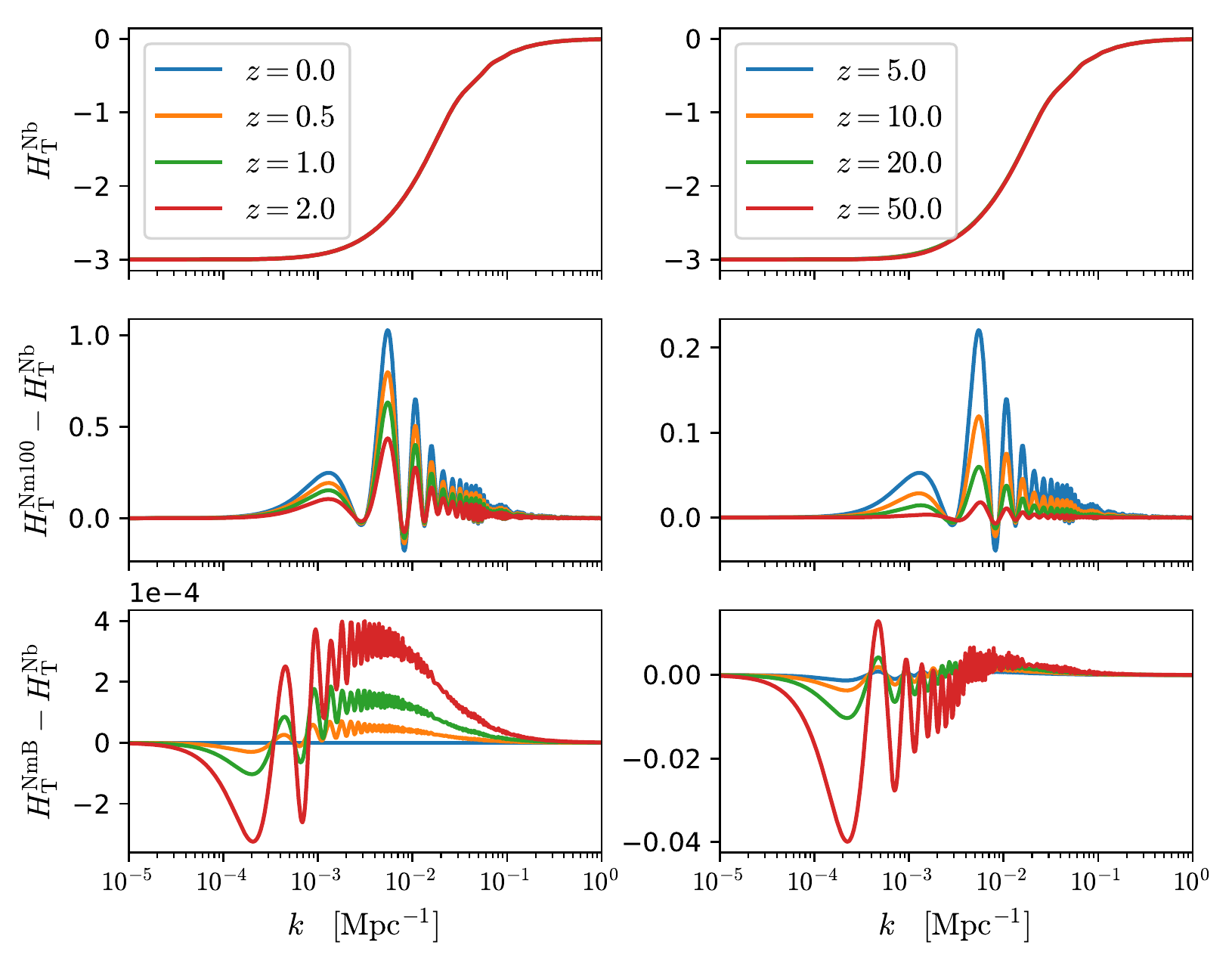}
  \caption{Comparison of the metric potentials, $H_{\rm T}(k)$, in the backward and the forward Newtonian motion gauge approach in a standard $\Lambda$CDM cosmology. \emph{Top row:} $H_{\rm T}^\text{Nb}=3\zeta$ in the N-body gauge. \emph{Centre row:} The deviation from the N-body gauge metric potential of $H_{\rm T}^{\text{Nm}100}$ in the ``forwards approach'' when the Newtonian motion gauge is initiated at $z=100$. The scale dependence is fixed by the early radiation and the deviation then grows with $D_+$.  \emph{Bottom row:} The deviation from the N-body gauge metric potential of $H_{\rm T}^{\text{NmB}}$ in the ``backscaling approach'', where the y-axis in the bottom left panel is rescaled by a factor of $10000$. For the observationally relevant redshifts the difference is below $10^{-3}$ on all scales, and the difference is always below $0.01$ for redshifts $z<20$. All modes are normalised to $\zeta=-1$ at super-horizon scales according to the \CLASS{}-convention.}
  \label{fig:figs_H_T_transfer}
\end{figure}

We have implemented these equations in \CLASS{} and present our results for a standard $\Lambda$CDM cosmology in figure~\ref{fig:figs_H_T_decaying2}. The metric potential $H_{\rm T}$ stays close to the N-body gauge value for the entire late-time evolution, while it changes significantly at the early times. This is opposed to the ``forwards approach'' discussed in subsection~\ref{forwards} and presented in our previous paper \cite{Fidler:2016tir}; this case is shown in the second panel of the figure. In figure~\ref{fig:figs_H_T_transfer} we show the same quantities but now as function of $k$ for eight different redshifts. The ``backscaling method'' evidently works extremely well for the analysed redshifts and scales. The corrections are still negligible at $z=20$, but quickly become relevant at earlier times. 

Our results can be understood in the following way. In the forward Newtonian motion gauges we start on the N-body gauge, utilising the N-body gauge initial displacements. However, due to the impact of radiation we are pushed away from the N-body gauge.
By contrast, in the backwards approach we start away from the N-body gauge, using different initial displacements. Then, due to the impact of radiation, we are pushed exactly onto the N-body gauge. As the N-body gauge is already a Newtonian motion gauge in the absence of radiation, we simply stay on that space-time for the remaining cosmic evolution in the standard $\Lambda$CDM cosmology. 

The relativistic initial matter displacements in the backwards gauge could be directly obtained from a linear Boltzmann code. Alternatively, we know that in the backwards Newtonian motion gauge the present day linear Newtonian density agrees with the linear N-body gauge density, since the volume deformation $H_{\rm L}$ is vanishing at that time, and we can scale back the present day Newtonian density to the initial time using $D_+$, even in the presence of radiation in the early Universe, 
\be
\delta^{\rm N}(\tau_{\rm ini}) =  D_+(\tau_{\rm ini}) \,\delta^{\rm Nb} (\tau_{\rm final}) \,.
\ee 
This relation corresponds exactly to the computation of the initial conditions in the 
conventional backscaling method~(\ref{eq:backIcs}), only that instead of the synchronous gauge density we use the N-body gauge at the present time. Note that the density in the synchronous gauge and N-body gauge are almost identical at the present time in a standard cosmology due to the absence of radiation.

This backwards gauge is not the only possible Newtonian motion gauge fixing. However it is the unique choice corresponding to a simple interpretation in terms of the standard back-scaling approach to set initial conditions. Only the choice of $C_{\pm}^{H_{\rm T}} = 0$ forces a constant metric potential $H_{\rm T} = 3\zeta$ in the absence of radiation. In addition the identification of the relativistic density and the simulation density at the final time requires a vanishing volume deformation $H_{\rm L} = 0$. These two conditions together uniquely select the N-body gauge as the metric at the present time. This means that any other gauge fixing will lead to a more complex dynamics for the relativistic metric potentials. The N-body gauge is thus uniquely suited for the ``backscaling approach''. The commonly used synchronous gauge density is suitable for backscaling only because it is identical to the N-body gauge density in the absence of radiation.

By gauge fixing the Newtonian motion gauge to the N-body gauge at the final time, we have provided the first fully relativistic interpretation of the backscaling method. We have explained why it is crucial to utilise the N-body gauge density at the present time and have constructed the backwards Newtonian motion gauge in which conventional N-body simulations can be interpreted and observables can be constructed in agreement with general relativity at first order.

\subsection{Summary of the relativistic backscaling prescription} 

We briefly summarise our backscaling prescription that enables a consistent relativistic interpretation of Newtonian simulations in a $\Lambda$CDM Universe:
\begin{itemize}
\item Using a linear relativistic Boltzmann code, generate the present day matter power spectrum in the N-body gauge. (For practical purposes the power spectrum in synchronous gauge is a good approximation and almost identical the one in N-body gauge at the present time). 
\item Scale the power spectrum back to the time of initialisation using the linear Newtonian growing mode, $D_+$ in eq.~(\ref{eq:backIcs}), consistently including radiation in the background.
\item Compute initial displacements from this power spectrum and utilise the logarithmic derivative of the growing mode, $f$, to set the initial velocities.
\item Evolve these initial conditions under Newtonian gravity to the present time, using a Newtonian N-body simulation containing pure pressureless dust, while including radiation in the background evolution. 
\item Interpret the output in the backwards Newtonian motion gauge coordinates, which can be computed using a linear Boltzmann code. For observables that focus on the late evolution this step simplifies as the corresponding Newtonian motion gauge is identical to the N-body gauge in the late Universe. Note that the N-body gauge metric and displacements are not identical to those in the synchronous gauge.\footnote{N-body gauge perturbations can be extracted from \CLASS{}. An implementation including Newtonian motion gauge perturbations is available on request.}
\end{itemize}

\section{Decaying dark matter}\label{sec:decay}

We have shown that the backwards Newtonian motion gauge is well suited for the analysis of standard $\Lambda$CDM models where the Universe is dominated at late-times by cold dark matter and a cosmological constant.  This is no longer true in the presence of radiation or other relativistic effects at late times, and we demonstrate this explicitly by investigating a decaying dark matter model~\cite{Audren:2014bca,Ichiki:2004vi,Adams:1998nr,DeLopeAmigo:2009dc,Gong:2008gi}, described in detail in appendix~\ref{app:ddm}.  

In the Newtonian picture, decaying dark matter primarily affects the cosmic evolution through its impact on the background.
A significant fraction of the dark matter density is transformed into decay radiation, changing the expansion of the Universe.  This effect can be partially modelled in a Newtonian simulation by reducing the mass of the dark matter particles as a function of the conformal time.   In this Newtonian limit, the decay is assumed to occur homogeneously.  

However, in the full relativistic theory, the impact of decaying dark matter is more complex. The decay radiation affects the gravitational forces acting on the matter particles and furthermore, the decay of dark matter is described by its local clock (comoving proper time). Observed in conformal time, this implies that the decay is faster or slower in different regions of space-time, which induces perturbations in the dark matter density. 

We will analyse this behaviour using the Newtonian motion gauge framework.
For the decaying fluid, described by the density contrast $\delta_{\rm dcdm}$ and the velocity $v_{\rm dcdm}$, we find the equations of motion, \begin{align}
\dot{\delta}_{\rm dcdm} + k v_{\rm dcdm} &= - 3 \dot{H}_{\rm L} + a \Gamma A  \,, \\
\left(\partial_{\tau} + \mathcal{H} \right)v_{\rm dcdm} &= - k (\Phi + \gamma) \,.
\end{align}
Here, the metric potential $A$, defined in \eqref{metric-potentials}, describes the inhomogeneous decay of the matter density. 
In addition we add a full Boltzmann hierarchy for the induced decay radiation contributing to $\delta_{\rm rad}$, $v_{\rm rad}$, $p$ and $\Pi$.
We define a Newtonian motion gauge in the usual way by enforcing $\Phi +\gamma = \Phi^{\rm N}$.

Employing the continuity equations we obtain the relation between the Newtonian density and the relativistic density,
\be \label{eq:deltaNdiff}
\delta_{\rm dcdm}^{\rm N} - \delta_{\rm dcdm}  = 3 H_{\rm L}  - \Gamma \int \dd\tau \, a A \,,
\ee
The Newtonian matter density, $\delta_{\rm dcdm}^{\rm N}$, is a pure coordinate density and does not include the impact of the relativistic volume deformation. The relativistic density on the other hand is based on the same particle positions (as we are in a Newtonian motion gauge), but evaluates them on the non-trivial space-time, inducing the correction of $3 H_{\rm L}$.
The second term describes the generation of dark matter perturbations by particle decays in regions where cosmic and dark matter times do not coincide.

This latter term is not implemented in the Newtonian simulation, but should not be ignored.  In particular, this contribution can be included in the relativistic density and added to the output of the Newtonian simulation as a post-processing. 
To do this, we define the density contrast $\xi$, describing the fraction of mass lost by matter particles based on their spatial position,
\be
\label{eq:xi}
\xi = - \Gamma \int \limits_{\tau_{\rm ini}}^{\tau} \dd\tau  \,a A.
\ee
where we have chosen to set $\xi = 0$ initially, corresponding to no decays of dark matter prior to the initial time.

As described in section~\ref{sec:NMrecap}, the Newtonian motion gauge condition is enforced by the evolution of $H_{\rm T}$, and its evolution in this model is determined by  
\begin{align}
\left(\partial_{\tau} +  \mathcal{H}\right) & \dot{H}_{\rm T} -  4 \pi G a^2 \bar \rho_{\rm dcdm}\left( H_{\rm T} -3\zeta \right)=  \nonumber \\
& 4 \pi G a^2 \left( \bar \rho_{\rm other}\delta_{\rm other}   +  3 \mathcal{H}k^{-1} ( \bar \rho + \bar p)_{\rm other}(v-B) - \bar \rho_{\rm dcdm} \xi + 2 \bar p \Pi\right).
\end{align}

We see that the decay of dark matter introduces an extra source term proportional to the decay fraction $\xi$. In addition, the late-time presence of the decay radiation introduces non-vanishing terms: $\delta_{\rm other}$, $p_{\rm other}$ and $\Pi$. 
  
\subsection{Numerical results}

We perform an explicit calculation of the Newtonian motion gauge space-time for the relativistic interpretation of Newtonian simulations including decaying dark matter. In this section we work in the comoving temporal gauge $B=v$. 
We have implemented the relevant equations in \CLASS{} and solve for the metric potentials and the decay fraction \eqref{eq:xi} in linear theory. We study two scenarios including dark matter that are compatible with current observations. In the first scenario, called ``DCDM'' in our plots, the entirety of the dark matter is decaying, while in the second scenario, called ``CDM+DCDM'', dark matter is made up by two species of which only one is decaying. For further details on the models we refer to appendix~\ref{app:ddm}. 

\begin{figure}[tb]
  \centering
    \includegraphics[width=0.95\textwidth]{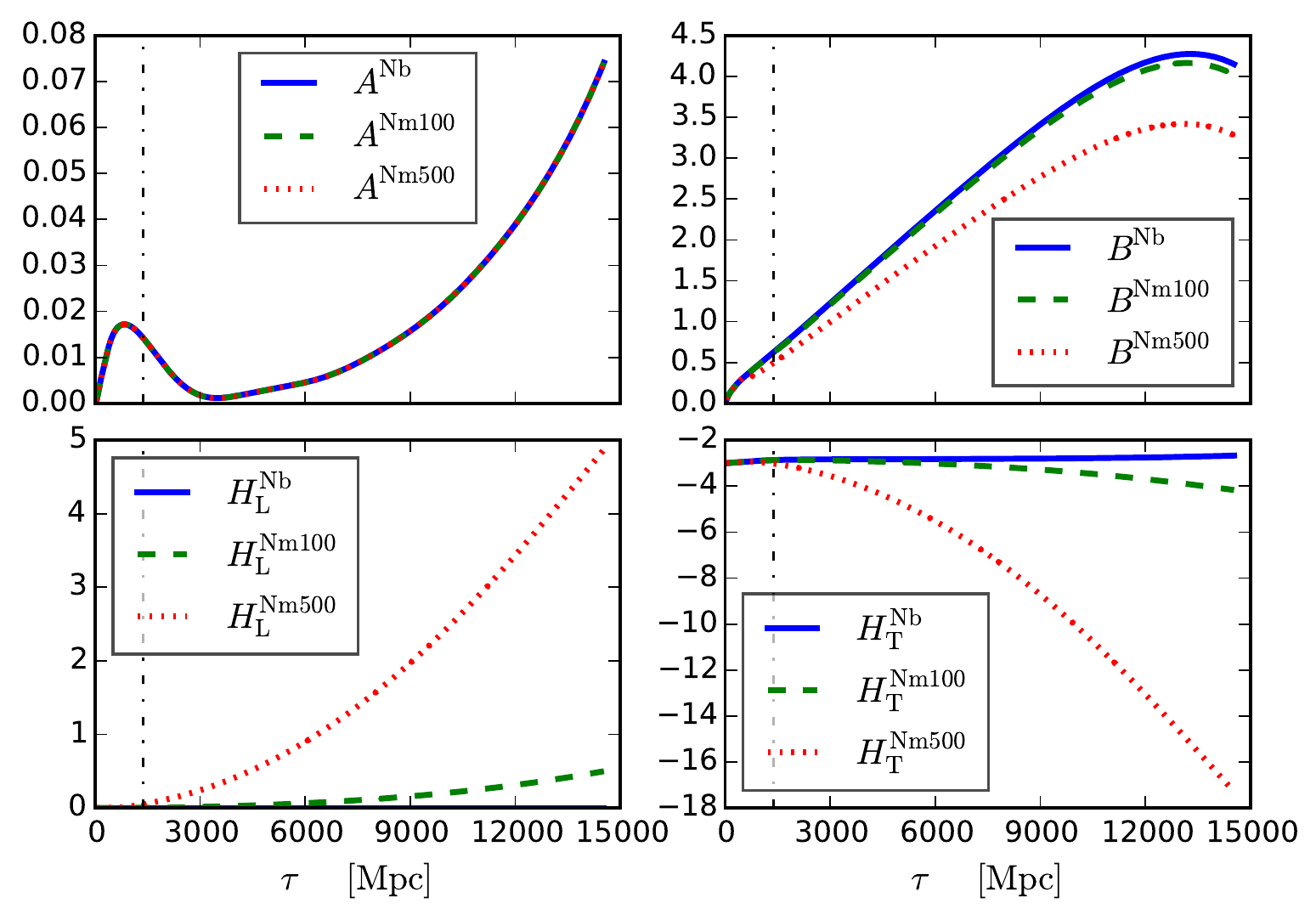}
  \caption{Metric potentials in the forward Newtonian motion gauge approach in a cosmology where dark matter decays with a lifetime of $\Gamma_\text{dcdm}^{-1}=160\text{Gyr}$, leading to the presence of radiation at late times. The curves represent different initialisation times of the forward Newtonian motion gauge ($z=100$ and $z=500$). Here we show the time evolution of all four metric potentials for the mode $k=0.002 \text{Mpc}^{-1}$. All the forward-Nm gauge metric potentials remain perturbative. The mode is normalised to $H_{\rm T}=3\zeta=-1$ at super-horizon scale.}
  \label{fig:figs_DecayMetricPotentialsk001}
\end{figure}

Our results for the forward Newtonian motion gauge are summarised in fig.~\ref{fig:figs_DecayMetricPotentialsk001}, where we show a typical initialisation redshift of $z=100$ and in addition a very early initialisation at $z=500$ where the metric in comparison is affected mostly by the early radiation effects.
The metric potential $A$ vanishes in the absence of radiation. In the presence of dark matter decays we find that $A$ is growing due to the induced decay radiation. We see that $H_{\rm T}$ starts on the $N$-body gauge value, but is quickly driven away from the N-body gauge by residual radiation at early times. The presence of the decay radiation further enhances this at late times. At the same time the volume deformation $H_{\rm L}$, which vanishes in the $N$-body gauge, is induced. 
Nonetheless the metric perturbations remain of the same order of magnitude as the comoving curvature perturbation $\zeta$, suggesting a perturbative approach to calculating the space-time remains valid.

\begin{figure}[tb]
  \centering
    \includegraphics[width=0.95\textwidth]{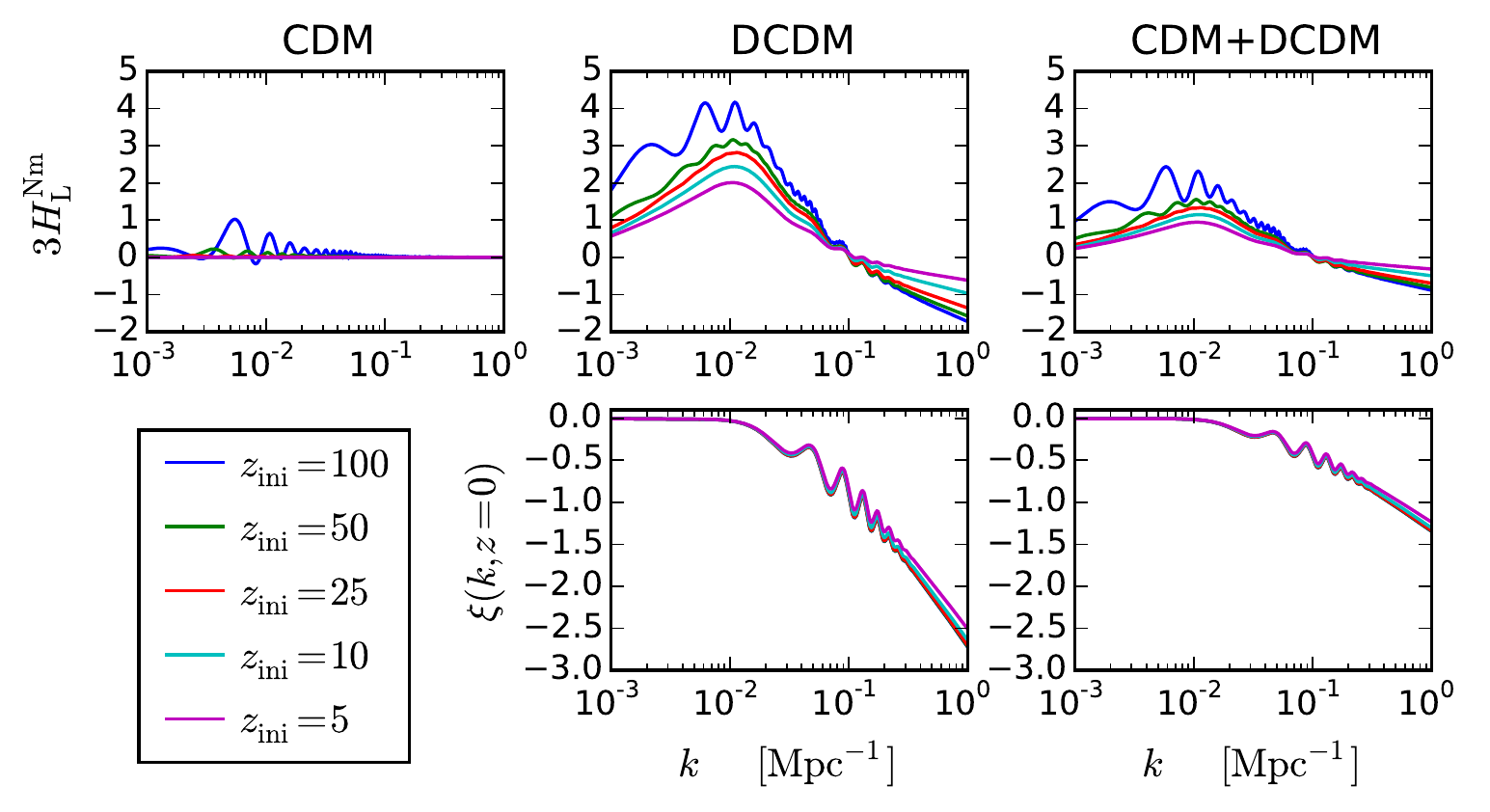}
  \caption{We plot the two relativistic corrections to the Newtonian matter density, the volume deformation~$3H_{\rm L}$ (top row) and the decay fraction $\xi$ in Eq.~\eqref{eq:xi}, due to inhomogeneous dark matter decay (bottom row). The various lines correspond to different initialisation times for the Newtonian motion gauge in the ``forwards approach''. The top left panel shows the CDM case without decaying dark matter, hence $\xi=0$ in this case. The deviation of $3H_{\rm L}$ from zero is due to residual radiation at early times, so initialising later reduces $3H_{\rm L}$. For the two models including decaying dark matter (as described in the appendix), $3H_{\rm L}$ is mostly dominated by the late-time radiation from the dark matter decay, so it depends rather weakly on the initialisation redshift. The $\xi$ correction has almost no dependence on the time of initialisation as long as it is before $z=5$ since most of the dark matter decays happen later.}
  \label{fig:figs_L_and_M_decaying}
\end{figure}

A Newtonian simulation with decaying dark matter can thus be given a relativistic interpretation, by embedding it in this particular space-time and changing the final particle masses according to the decay fraction $\xi$. In fig.~\ref{fig:figs_L_and_M_decaying} we show the resulting difference between the simulation density and the relativistic density by comparing the volume deformation $3H_{\rm L}$ and the decay fraction $\xi$, the two corrections appearing in equation~\eqref{eq:deltaNdiff}. The left panel describes a standard $\Lambda$CDM cosmology without decays, and thus $\xi = 0$. The volume deformation at the present time, $3H_{\rm L}$, is vanishing if the simulation is initiated at sufficiently late times when radiation may be neglected. The right panels show the results in our two models including decaying dark matter. The decay fraction, $\xi$, is the dominant effect on the smaller scales and does not depend significantly on the initialisation time.
The volume deformation, $3H_{\rm L}$, contains the oscillations from early residual radiation, in agreement with the non-decaying case. On top of this, there is a smooth contribution from late decay radiation and, on the small scales, the impact of the inhomogeneous decay fraction $\xi$. 

\begin{figure}[tb]
  \centering
    \includegraphics[width=0.95\textwidth]{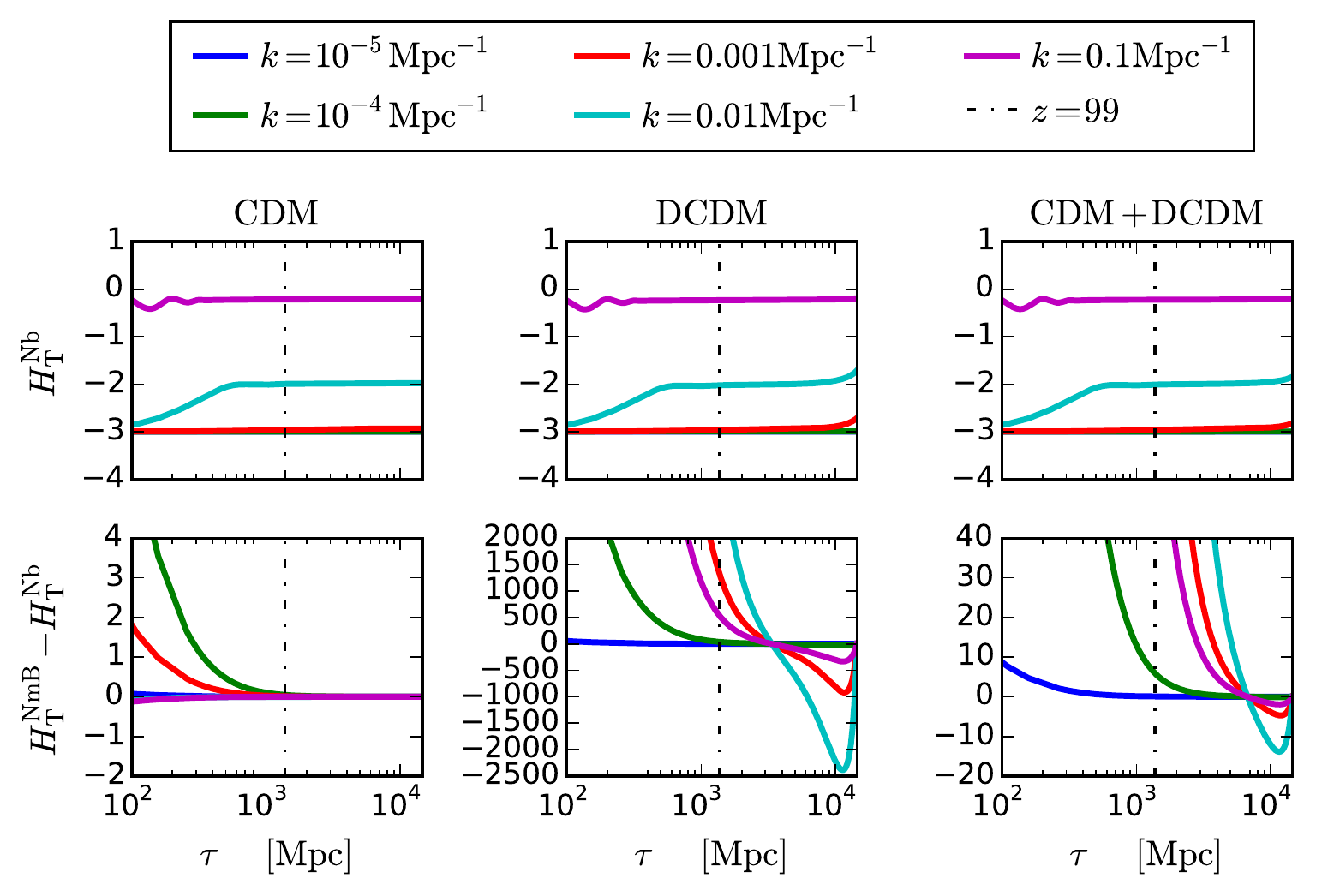}
  \caption{\emph{Top row:} The N-body gauge metric potential, $H_{\rm T}^\text{Nb}$, for the three scenarios, standard cold dark matter (CDM), decaying dark matter (DCDM) and mixed CDM+DCDM. \emph{Bottom row:} The difference between the metric potential in the backwards Newtonian gauge, $H_{\rm T}^\text{NmB}$, and the $N$-body gauge potential,  $H_{\rm T}^\text{Nb}$. In the CDM case all modes quickly converge on the $N$-body gauge value as shown previously. However, for the decaying dark matter scenarios the perturbative description breaks down in the ``backscaling approach''; the gauge potentials become large at early times and cannot be reliably computed in a first-order Boltzmann code.}
  \label{fig:figs_H_T_decaying}
\end{figure}

We also analyse our models in the backwards Newtonian gauge.
The resulting metric is plotted in fig.~\ref{fig:figs_H_T_decaying}, where the late-time presence of radiation induces evolution in the metric over the whole lifetime of the Universe.
In order to finish on the $N$-body gauge space-time, the metric potentials have to start at very large initial values and only approach the simple $N$-body gauge at the final time. 
It is thus no longer consistent to analyse the metric potentials in a perturbative approach
and the backscaling procedure should be avoided.

By comparison, the metric perturbations in the ``forwards approach'' remain small at all times. It is particularly efficient for including late-time sources of radiation, which only affect the metric evolution for a short period of time. Early radiation can also be included in this framework so long as the sources are relatively weak. In contrast, in the ``backscaling approach'' the early metric potentials are affected by all sources that will be relevant in the future, making late-time sources problematic, while it is more reliable for early sources. 

We conclude that backscaling is well-suited to including early, residual radiation in the Universe, but late-time relativistic effects such as decaying dark matter should be avoided. For these the forward Newtonian motion approach is better suited as the metric perturbations stay small. 

\section{Conclusions}\label{sec:conclude}

N-body simulations are necessarily performed using only an approximate physical model; they are intrinsically Newtonian rather than fully relativistic and they do not include all fluid species, including radiation at most through its effect on the background expansion. In addition, they treat the baryons and dark matter particles as a single pressureless fluid, neglecting the fact that the coupling to radiation in the early Universe leads to different initial positions and velocities~\cite{Tseliakhovich:2010bj}. 

Some of these deficiencies are addressed by the 
``backscaling approach'', which compensates for the lack of radiation and mixed baryon+CDM initial conditions 
by starting with a set of \emph{fictitious} initial displacements designed to reproduce the linear-scale synchronous gauge total matter density at the present.  However, a number of questions remain regarding this approach, in particular how the late-time non-linear evolution is affected by missing elements, such as radiation perturbations and anisotropic stress arising from matter on non-linear scales. 
Moreover, there is no relativistic interpretation of gauge-dependent quantities such as displacements and densities in a purely Newtonian framework.

We have addressed these questions by constructing a consistent relativistic treatment based on the Newtonian motion gauge approach~\cite{Fidler:2016tir}, 
where we construct a relativistic space-time where the matter follows the Newtonian trajectories calculated in the N-body code.
By matching the linear matter density to the relativistic boundary conditions at the final time, we reproduce the ``backscaling approach'' in the matter sector.  The physical interpretation of these results requires also the metric potentials in the Newtonian motion gauge, in particular the trace-free spatial metric perturbation $H_{\rm T}$, whose evolution consistently accounts for the presence of inhomogeneous radiation so long as it remains perturbative.  

These solutions are particularly simple in the radiation-free limit, where the Newtonian motion gauge reduces to the N-body gauge and the metric perturbation is constant, $H_{\rm T} = 3 \zeta$, where $\zeta$ is the 
comoving curvature perturbation.  The presence of radiation in the early Universe causes the N-body gauge and Newtonian motion gauges to diverge, but by matching them at the present, we have shown that they coincide for most of the period where non-linear structures are growing.  

Using Newtonian motion gauges, we specify the underlying space-time and 
show that in standard $\Lambda$CDM cosmologies the metric perturbations remain small, and can consistently be solved using linear Boltzmann codes.
Newtonian motion gauges provide 
the fully relativistic space-time, enabling, for example, the use of relativistic ray-tracing.

In theories beyond the standard model backscaling must be performed with great care. The space-time may not have a simple late-time limit and the metric potentials may become large.
We have presented an example including decaying dark matter and conclude that backscaling is not an appropriate method to generate initial conditions for simulating such cosmologies. Instead the forward Newtonian motion approach may be employed and remains
well defined. In general we conclude that modifications to the late-time dynamics typically invalidate backscaling. For these models only a direct analysis in the Newtonian motion gauge framework can show whether backscaling can be used.

Backscaling could also be applied for simulating two distinct baryon and cold dark matter fluids. These fluids follow an identical velocity distribution at the present time. Employing backscaling we thus obtain initial conditions for a single fluid Universe that is designed to reproduce the present day power spectrum when fully including baryons and cold dark matter.
This approach is discussed in \cite{Valkenburg:2016xek}, comparing the backscaling method to setting initial conditions for both baryons and cold dark matter in the N-body gauge, however neglecting the impact of radiation. 
Employing the Newtonian motion gauges, we can use exact initial conditions for both species in the ``forwards approach'' and thus fully include the baryon physics in the simulation. With backscaling, however, the decreasing difference between the baryon and cold dark matter velocities and densities resemble a decaying mode and may cause the metric potentials to diverge, comparable to the dynamics we observe in the decaying dark matter case.
Thus, while a ``forwards approach'' may be employed, detailed follow-up studies are required to investigate the performance of backscaling and baryons, which we leave for future work.

Another highly important application for backscaling 
is generating initial conditions of N-body simulations in the presence of massive neutrinos (see \cite{Zennaro:2016nqo} for a Newtonian perspective), or modified gravity \cite{Valkenburg:2015dsa}. Massive neutrinos in particular resemble dark matter at late times, whilst they contribute to the radiation content at early times. In the language of Newtonian motion gauges this represents a modification to the ``Newtonian'' dynamics at the early times. In a forthcoming study we will investigate which type of gauges are suitable for this problem.

\acknowledgments

CF is supported by the Wallonia-Brussels Federation grant ARC11/15-040 and the Belgian Federal Office for Science, Technical \& Cultural Affairs through the Interuniversity Attraction Pole P7/37.
TT, RC, KK, and DW are supported by the UK Science and Technologies Facilities Council grants ST/N000668/1. 
CR is supported by the DFG through the SFB-Transregio TRR33 ``The Dark Universe''. 
KK is supported by the European Research Council through grant 646702 (CosTesGrav).
\\[1.5cm]

%%%%%%%%%%%%%%%%%%%%%%%%%%
\appendix 

\section{Decaying dark matter}\label{app:ddm}

We consider two models, both with dark matter decay rate $\Gamma = 6~{\text{km}}~{\text{s}^{-1} \text{Mpc}^{-1}}$, which is the maximal allowed value if all of dark matter is decaying~\cite{Audren:2014bca,Ichiki:2004vi,Adams:1998nr,DeLopeAmigo:2009dc,Gong:2008gi}. It corresponds to a lifetime of $\Gamma^{-1} = 160~\text{Gyr}$. In the first model, denoted ``DCDM``, all dark matter is assumed to be decaying. In the second model, denoted ``CDM+DCDM'', half of the dark matter is assumed stable while the other half is decaying. 
The dark matter linear equations of motion read
\begin{align}
\dot{\delta}_{\rm cdm} + k v_{\rm cdm} &= - 3 \dot{H}_{\rm L} \,,\\
\dot{\delta}_{\rm dcdm} + k v_{\rm dcdm} &= - 3 \dot{H}_{\rm L} - a \Gamma A \,, \\
\left(\partial_{\tau} + \mathcal{H} \right)v_{\rm cdm} &= - k (\Phi + \gamma) \,,\\
\left(\partial_{\tau} + \mathcal{H} \right)v_{\rm dcdm} &= - k (\Phi + \gamma) \,, 
\end{align}
where the metric potential $A$ is the perturbation in the lapse function and thus leads to an inhomogeneous decay at given coordinate time.

The dynamical equations for the velocities are identical, and given the same initial velocity distribution the velocities of both types of dark matter stay identical at first order. This also allows us to define a Newtonian motion gauge in which both species follow Newtonian trajectories simultaneously,
\be
v_{\rm cdm} = v_{\rm dcdm} 
= v_{\rm m} \,.
\ee 
We will define a combined fluid describing the mix of both dark matter types: 
\be
 \bar \rho_{\rm m}\delta_{\rm m} = \bar \rho_{\rm cdm}\delta_{\rm cdm}+ \bar \rho_{\rm dcdm}\delta_{\rm dcdm} \,.
\ee
For this combined fluid we find the dynamical equations
\begin{align}
\dot{\delta}_{\rm m} + k v_{\rm m} &= - 3 \dot{H}_{\rm L} + a \Gamma \frac{\bar \rho_{\rm dcdm}}{\bar \rho_{\rm m}}(\delta_{\rm m}-\delta_{\rm dcdm} + A)  \,, \\
\left( \partial_{\tau} + \mathcal{H} \right)v_{\rm m} &= - k (\Phi + \gamma) \,. 
\end{align}
We relate the density of the combined fluid to the Newtonian density by comparing the continuity equations:
\be
\delta_{\rm N} - \delta_{\rm m} - 3 H_{\rm L}  = \xi
\ee
with the decay fraction
\be
\xi = - \Gamma \int \limits_{\tau_{\rm ini}}^{\tau} \dd\tau a \frac{\bar \rho_{\rm dcdm}}{\bar \rho_{\rm m}}(\delta_{\rm m}-\delta_{\rm dcdm} + A)
\ee
Numerical results for the two relativistic contributions to the total matter density, $H_{\rm L}$ and $\xi$, are shown in figure~\ref{fig:figs_L_and_M_decaying} for both the decaying dark matter and the mixed CDM plus decaying dark matter models. In figure~\ref{fig:figs_H_T_decaying} we show the metric potentials in the forwards and backwards Newtonian motion gauges for these scenarios.

\bibliographystyle{JHEP}
\bibliography{references}

\end{document}